\documentclass[aps,superscriptaddress,floats,twocolumn,epsf,prl]
{revtex4-2} 
\usepackage{amsmath,amsfonts}
\usepackage{mathtools}
\usepackage{eufrak}
\usepackage{multirow}
\usepackage{textcomp}
\usepackage{units} 
\usepackage{xfrac}
\usepackage{orcidlink}

\usepackage{graphicx}

\usepackage{MnSymbol}
\usepackage{nicefrac}
\usepackage{braket}
\usepackage{array}
\usepackage{bm}
\usepackage{hyperref}
\usepackage{changes}
\usepackage[normalem]{ulem}

\usepackage{hyperref}
\hypersetup{colorlinks=true,breaklinks,linkcolor=blue,urlcolor=blue,citecolor=blue}

\begin{document}
\title{Magnetic quantum criticality: The role of the Fermi surface geometry}
\author{D.~R.~Fus\orcidlink{0009-0007-4425-6728}}
\affiliation{Institute of Solid State Physics, TU Wien, 1040 Vienna, Austria}
\author{S.~Adler\orcidlink{0000-0002-0736-0661}}
\affiliation{Institute of Solid State Physics, TU Wien, 1040 Vienna, Austria}
\author{M.~O.~Malcolms\orcidlink{0000-0002-5502-8583}}
\affiliation{Max Planck Institute for Solid State Research, Heisenbergstra{\ss}e 1, 70569 Stuttgart, Germany}
\author{A.~Vock} 
\affiliation{Institute of Solid State Physics, TU Wien, 1040 Vienna, Austria}
\author{K. Held\orcidlink{0000-0001-5984-8549}}  
\affiliation{Institute of Solid State Physics, TU Wien, 1040 Vienna, Austria}
\author{A.~A.~Katanin\orcidlink{0000-0003-1574-657X}}
\affiliation{Center for Photonics and 2D Materials, Moscow Institute of Physics and Technology, Institutsky lane 9, Dolgoprudny, 141700, Moscow region, Russia}
\affiliation{M. N. Mikheev Institute of Metal Physics, Kovalevskaya Street 18, 620219 Ekaterinburg, Russia}
\author{T.~Sch\"afer\orcidlink{0000-0002-1105-5619}}
\affiliation{Max Planck Institute for Solid State Research, Heisenbergstra{\ss}e 1, 70569 Stuttgart, Germany}
\affiliation{Dipartimento di Fisica, Università di Trieste, Strada Costiera 11, I-34151 Trieste, Italy}
\author{A.~Toschi\orcidlink{0000-0001-5669-3377}}
\affiliation{Institute of Solid State Physics, TU Wien, 1040 Vienna, Austria}


\begin{abstract}
We investigate magnetic quantum phase-transitions in bulk correlated metals. 
To this end, we focus on the Hubbard model on different cubic lattices as a function of temperature and electronic density, determining the relevant regimes around its quantum magnetic transition -- classical, quantum critical, and quantum disordered-- as well as the corresponding (thermal/non-thermal) quantum critical exponents. Our numerical results, based on dynamical mean-field theory, together with supporting analytical derivations, rigorously demonstrate \emph{how} and \emph{why} the presence of different kinds of Kohn anomalies on the underlying Fermi surface (i) drives the quantum critical behavior above the quantum critical point and (ii) shapes the whole phase diagram around it. Our findings highlight the importance of an explicit inclusion of such Fermi-surface geometrical properties into the universality class definition for magnetic \emph{quantum} phase-transitions in correlated metals. 

\end{abstract}

\maketitle

{\sl Introduction.} Although the phase diagrams of correlated materials are particularly rich of quantum critical points (QCPs) and associated zero-temperature phase-transitions, a full comprehension of their critical behavior still poses substantial hurdles to the many-electron theory \cite{Sachdev1999,Coleman2005,Kopp2005,Sachdev2011}. 
This is particularly true in the relevant case of correlated metals, where several issues hinder 
a standard interpretation 
of their quantum critical properties \cite{Loehneysen2007,Sachdev2010,Sachdev2011,Brando2016}. The main reasons behind these difficulties are the underlying presence of the Fermi surface (FS) and of corresponding low-energy excitations, as well as the multifaceted \cite{Qi2001,Byczuk2007,Held13,Gleis2023}, possibly nonperturbative \cite{Kozik2015,Reitner2020,vanLoon2020,Chalupa2021,Adler2024,Kowalski2024,Moghadas2026}, effects of poorly screened electronic interactions.

Within the conventional Hertz-Millis-Moriya (HMM \cite{Hertz1976,Millis1993,Dzyaloshinskii1976,Moriya1973}) description of quantum criticality, the dynamical effects
relevant for the electronic system considered are effectively included in a low-energy bosonic action, deriving the corresponding value of the dynamical critical exponent $z$. This encodes the effects of long temporal fluctuations for the specific  phase-transition of interest. In particular, $z$ in the HMM can differ from the value of $1$ defining the simplest realization of the quantum-classical mapping (valid, e.g., for the Ising model \cite{Sachdev2011}), improving the description of quantum criticality for the systems considered.
Nonetheless, the effective bosonic HMM-description may fail to capture relevant corrections to the quantum critical behavior arising from fermionic excitations and their (possibly not negligible) interactions.

\begin{figure*}[th!]
 \centering
 \includegraphics[scale=0.95, trim=0.3cm 0mm 0cm 0cm]{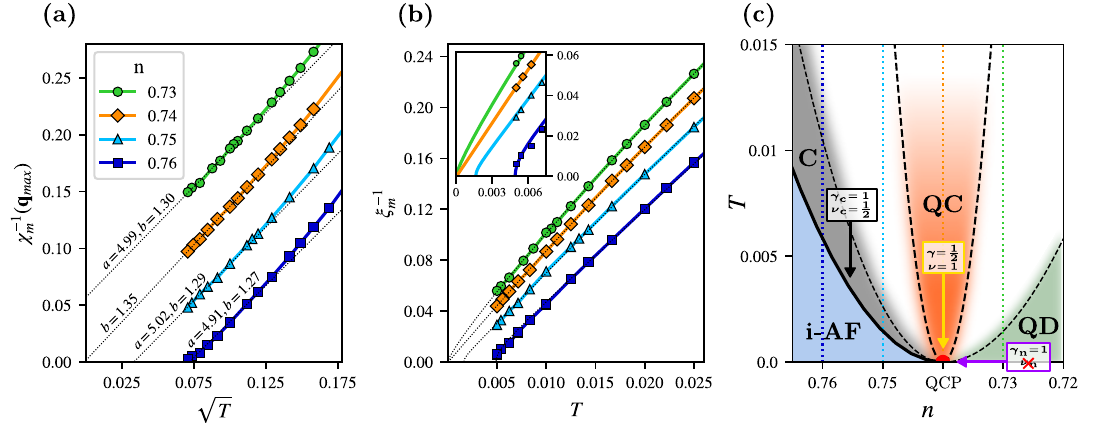}
\vspace{-5mm} \caption{(a,b) Inverse of the maximum static magnetic susceptibility $\chi^{-1}(\mathbf{q}_\text{max})$ and of the corresponding magnetic correlation length $\xi_m^{-1}$ of the DMFT solution of the 3D Hubbard model on a simple cubic lattice, plotted, respectively, as a function of $\sqrt{T}$ and $T$ for different values of the electronic density $n$.  Fitting parameters of $\chi_m^{-1}$  to Eq.~(\ref{eq:Invchi})  are reported in panel (a); a low-$T$ zoom of the fits of $\xi^{-1}_m$ is shown in the inset of (b). (c) Phase-diagram around the QCP, summarizing the obtained results.}
\label{fig:Fig1}
\end{figure*}

In this respect, previous studies based on the dynamical vertex approximation \cite{Toschi2007}, a non-local diagrammatic extension \cite{Rohringer2018} of dynamical mean-field theory (DMFT \cite{Metzner1989,Georges1992a,Georges1996}), 
have pointed out \cite{Schaefer2017,Stepanenko2017} the possible role played by the 
  underlying Fermi surface (FS) in affecting the magnetic response of the quantum critical regime. 
 
Indeed, Refs.~\cite{Schaefer2017,Stepanenko2017} observed, 
how the presence of Kohn points (KPs) on a three-dimensional (3D) FS (i.e., pairs of points coupled by the momentum of the dominant fluctuations, whose Fermi velocities are antiparallel) could trigger non-analyticities in the $T \! \rightarrow \! 0$ behavior of the physical susceptibilities, appearing \emph{in addition} to those normally associated to the quantum critical fluctuations. 
While effects of KPs in 2D were recently studied in \cite{Holder2012,Holder2014,Sykora2018,Halbinger2019,Sykora2021,Debbeler2023,Debbeler2024},  KPs in 3D appeared able to alter, albeit \emph{only} in the quantum critical regime, 
the temperature dependence of the magnetic response $\chi_m \propto T^{-\gamma}$ and of the associated magnetic correlation length $\xi_m \propto  T^{-\nu}$, as well as the  Fisher relation \cite{Fisher1964} linking the corresponding  (thermal) quantum critical exponents $\gamma$ and $\nu$. In the considered case of a 3D FS with \emph{lines} of Kohn anomalies, the Fisher relation for magnetic quantum phase-transitions in 3D ($\gamma \! = \! 2 \nu$) \footnote{While the Fisher relation generally reads $\gamma \! = \! (2 \! -\! \eta) \nu$ \cite{Binney2002,Goldenfeld1992}, we recall that $\eta$ is already much smaller w.r.t.~the other exponents for finite-$T$ magnetic transitions in 3D and gets suppressed by the additional dynamical fluctuations in the quantum critical regime.} was even apparently ``reverted'' to $\gamma \! = \! \frac{1}{2} \nu$, namely with $\gamma \! = \! \frac 12, \, \nu \! = \! 1$ \cite{Schaefer2017,Stepanenko2017}. This differs from both the corresponding HMM values  $\gamma \! = \! 2\nu= \! \frac 32$ (with $z\!= \!2$) and the Sommerfeld-MF ones $\gamma \! = \! 2\nu \! = \!2$. Such large modifications  are hardly ascribable to correspondingly large values of the anomalous exponent $\eta$, because, for the \emph{on-site} (i.e.~\emph{short-range}) interactions considered here, a \emph{full suppression} ($\eta \! = \!0$) is actually expected, due to the high effective dimensionality ($d_{\rm eff} \! =d \! +\!z$) of magnetic quantum phase-transitions in 3D.

{\sl Previous Studies vs.~our Strategy.} The huge numerical effort of the diagrammatic approach \cite{Toschi2007} used to study quantum criticality in \cite{Schaefer2017} and the additional approximations made  \cite{Katanin2009,Rohringer2016} prevented, however, to grasp
the general picture of the magnetic quantum phase-transition considered and its several, important implications. 
In particular, previous studies could \emph{not} assess to what extent KP-driven singularities may affect (i) the classical transition and (ii) the quantum disordered regime around the QCP,  whether they also control (iii) the non-thermal quantum critical exponents, and, finally, (iv) why the Fisher relation appears to be abruptly modified \emph{only} for the quantum critical exponents.
Aiming at clarifying \emph{all} fundamental aspects of this problem and their surprising interrelations, we present here a numerical study based on DMFT, supported by rigorous analytical derivations. DMFT enables
a much higher numerical precision than previous analyses in
determining the $T \! \rightarrow \!  0$ behavior of the magnetic properties in the relevant parameter region around the QCP. The improved precision is indeed essential for our scope,  paving the way to a \emph{clear-cut analytical} understanding of both thermal and non-thermal behaviors of the physical quantities $\chi_m$ and $\xi_m$, as well as the determination of the associated (quantum) critical exponents.  We are, thus, able to unveil the \emph{direct link} between the specific FS-properties of 3D correlated metals and the \emph{overall shape} of the phase-diagram associated to their magnetic QCP in all relevant regimes (classical, quantum critical, quantum disordered). Finally, we can precisely ascribe the modified Fisher relation for quantum critical exponents to an emergent $T\! \rightarrow 0$ singularity of the spin-stiffness for 3D correlated metals induced by the KP anomalies on their FS. As we will show, this singularity \emph{directly} reflects the \emph{non-exponential}, power-law spatial decay of magnetic correlations in the \emph{ground state} of our system. Remarkably, this intrinsically \emph{long-range} (LR) 
behavior of  the $T\!=\!0$ correlations, caused by the KPs, does not only affect the quantum critical description of the ground state, but  crucially reverberates into the (more easily accessible) finite-$T$ properties of the quantum critical regime. 

{\sl Model and Methods.} We consider the 3D Hubbard model (HM  \cite{Hubbard1963,Hubbard64,Kanamori63,Gutzwiller63,Qin2022,Arovas2022}) on a simple cubic lattice with  nearest-neighbor hopping $t$, whose Hamiltonian is given in the Supplemental Material~\cite{Suppl}. Throughout this work $t \equiv 1$ serves as unit of energy, the lattice constant $a\equiv 1$ as unit of length, 
and the interaction $U$ is fixed at an intermediate-to-large value of $U \simeq 10 t$ (precisely: $U=9.789t$, for which at half filling the N{\'e}el temperature is maximal in DMFT \cite{Rohringer2011}), while the density $n$ and the temperature $T$ are varied.

The HM is studied in its paramagnetic phase by means of DMFT, computing both one- and two-particle \cite{Rohringer2012} properties. 
We can thus calculate via matrix inversion the (static \footnote{Computing the static susceptibility, i.e., the one for zero transfer Matsubara frequency, allows to describe the onset of second-order phase-transitions at any finite $T$.}) momentum-dependent spin/magnetic susceptibility, $\chi_m({\bf q})$, using the Bethe-Salpeter equation (BSE) in the corresponding channel \cite{Georges1996}:
\begin{equation}
\chi_m({\bf q}) = T^2 \sum_{\nu \nu^\prime} \left[(\chi_{0,\bf q}^{\nu ^\prime })^{-1}\delta _{\nu \nu^\prime}-\Gamma_m^{\nu \nu^\prime} \right]^{-1}, 
\label{eq:BSE}
\end{equation}
where $\chi_{0,\bf q}^{\nu ^{\prime }}=-T^{-1} \sum_{\mathbf{k}}G_{\bf{k}%
,\nu ^{\prime }}G_{\bf{k}+\bf{q},\nu^\prime}
$ is the
particle-hole bubble, $G_{\mathbf{k},\nu }=[i\nu -\varepsilon_{\bf k} + \mu
-\Sigma_\nu]^{-1}$ is the  lattice Green's function of DMFT, i$\nu$ the fermionic Matsubara frequency, $\varepsilon_{\bf k}$ the energy-momentum dispersion of the 3D cubic lattice, $\mu$ the chemical potential, $\Sigma_\nu$ the (local) self-energy, and the matrix $\Gamma_{m}^{\nu \nu^\prime}$ the local irreducible vertex in the magnetic channel. $\Sigma_\nu$ and $\Gamma_{m}^{\nu \nu^\prime}$ are obtained by inverting the Dyson equation and the BSE of the auxiliary impurity model of the self-consistent DMFT solution. For further algorithmic/technical details see \cite{Suppl}.

{\sl Numerical results.} Our DMFT results are reported in Fig.~\ref{fig:Fig1}. Here, the inverse of the maximum of the static magnetic susceptibility $\chi^{-1}_m$, which is achieved for ($T-$dependent) incommensurate values \cite{Zadeh1997,Schaefer2017,Vilardi2018,Lenihan2022,Rampon2025} of the momentum ${\bf q}_{\rm max}(T)= (\pi,\pi, \bar{q}_z=\pi - \delta(T))$, as well as the inverse of the corresponding magnetic correlation length $\xi^{-1}_m$ are shown for four different values of the electronic density $n$ in the relevant regime for the quantum phase-transition of our interest. In particular, plotting $\chi^{-1}_m$ as a function of $\sqrt{T}$ highlights one of our main findings: In the density interval at and around the QCP considered, $\chi^{-1}_m$ displays a universal $\sqrt{T}$  behavior, i.e.,
\begin{equation}
\chi^{-1}_m({\bf q}_{\rm max}, T,n)= a(n_c -n) + b\sqrt{T},
\label{eq:Invchi}
\end{equation}
where $a,b$ are positive constants [whose fitted values are explicitly reported in Fig.~\ref{fig:Fig1}(a)] and $n_c$ is the quantum critical density, which defines the location of the QCP on the density axis. In fact, while the  $\sqrt{T}$-dependence of $\chi^{-1}_m$ is immediately recognizable from the parallel alignment of all low-$T$ data plot in Fig.~\ref{fig:Fig1}(a), the equal distance of datasets for the fixed density-difference steps of $\Delta n=0.01$  reflects the linear $n$-behavior of the first term in the r.h.s.~of Eq.~(\ref{eq:Invchi}).
More quantitatively, 
the vanishing $T\!\rightarrow\!0$ extrapolation of $\chi^{-1}_m$ identifies, to a good precision, the value of $n_c \simeq 0.74$. Hence, for $n\!>\!n_c$, a classical phase-transition from a paramagnetic state to an incommensurate antiferromagnetic ordering should be expected at the critical temperature $T_c$ given by
\begin{equation}
    \chi_m^{-1}({\bf q}_{\rm max},T,n) \overset{!}{=}0 \; \;  \stackrel{\text{Eq.}~(\ref{eq:Invchi})}{\Rightarrow} \; \; \, T_c \;   {=}  \,   \frac{a^2(n -n_c)^2}{b^2}.
    \label{eq:TNeel}
\end{equation}
This expression features a somewhat remarkable {\sl parabolic} behavior  of $T_c$ (instead of the ``standard" square-root one 
\cite{Sachdev1999}) as function of the density [cf. the phase diagram sketch in Fig.~\ref{fig:Fig1}(c)].

The numerical data of $\xi_m^{-1}(T,n)$ shown in Fig.~\ref{fig:Fig1}(b) have been obtained by fitting the momentum dependence of the magnetic susceptibility computed in Eq.~(\ref{eq:BSE}) w.r.t.~an expression  {\sl \`a la} Ornstein-Zernike (O.Z.) \cite{Ornstein1916,Altland2008} along the direction $(\pi,\pi, q_z)$ \cite{Schaefer2017,SchaeferThesis}: 
\begin{equation}
\chi^{-1}_m(q_z,T,n) = {\cal A}^{-1}[(q_z \! - \! \bar{q_z})^2 +\xi_m^{-2} + {\cal B}(q_z \!-\! \bar{q_z})^3],
\label{eq:OZfit}
\end{equation}
where ${\cal A}$, $\xi_m$, and ${\cal B}$ are fitted, and the last term accounts for the asymmetric behavior displayed by $\chi^{-1}_m(q_z,T,n)$ around $\bar{q_z}$ for an underlying incommensurate magnetic instability.
The corresponding numerical data in Fig.~\ref{fig:Fig1}(b) display a roughly {\sl linear} temperature dependence of $\xi^{-1}$ in the $T,n$-range considered. Contrary as for $\chi^{-1}_m(T)$, however, the distance between the dataset of $\xi^{-1}$ at different $n$ is not equally spaced, indicating a nonlinear dependence on $n-n_c$.

{\sl (Quantum) Criticality and Phase Diagram.} The general physical picture emerging from our DMFT calculations can be best understood starting from the clear-cut behavior of $\chi_m^{-1}$, precisely captured by Eq.~(\ref{eq:Invchi}). In fact, Eq.~(\ref{eq:Invchi}), whose functional form can be {\sl analytically} linked to the Kohn-line anomalies (s.~\cite{Suppl}), nicely encodes the information about both thermal (for $n \! = \! n_c$; $T \! \rightarrow \! 0$) and non-thermal (for $T \! =\! 0$; $n \! \rightarrow \! n_c$) quantum critical exponents, i.e., $\gamma$ and  $\gamma_n$, respectively. Here, $\gamma_n$ defines the $T \!= \! 0$-behavior of $\chi_m^{-1}$ in the quantum disordered regime w.r.t.~a non-thermal control parameter, i.e., $\chi^{-1}_m \propto (n_c \!  - \! n)^{\gamma_n}$ for $n \! < \! n_c$. In particular, setting $n\! = \! n_c$ directly yields the peculiar value of $\gamma \! = \! \frac 12$, 
demonstrating how the effects of Kohn-anomalies on the FS \cite{Schaefer2017,Stepanenko2017, Suppl}, encoded in the bubble term of Eq.~(\ref{eq:BSE}), get directly mirrored onto the quantum critical properties of DMFT.
At the same time, the evaluation of Eq.~(\ref{eq:Invchi}) at $T\!=\!0$ yields for the nonthermal quantum critical exponent the value of $\gamma_n= 1$, consistent with the results of the HMM theory for $d_\text{eff} = d +z > 4$, indicating that Kohn-anomalies do {\sl not} affect the quantum critical behavior of the maximal susceptibility at $T\!= \! 0$ (for $n \! < \! n_c$). 
On a different note, the overall $\sqrt{T}$ behavior of $\chi^{-1}_m$  might appear, at a first sight, incorrect for $n>n_c$, because, in the proximity of the classical phase-transition at $T_c$, one would expect, in DMFT, the mean-field value for the corresponding critical exponent, i.e., $\gamma_\text{cl} \! = \! 1$.  In fact, $\gamma_\text{cl}$ is indeed $1$, as one easily realizes by Taylor expanding Eq.~(\ref{eq:Invchi}) for $T \! \gtrsim \! T_c$ (and $n \! \gtrsim \!  n_c$), which yields $\chi_m^{-1}(T \! \gtrsim \!  T_c) =  \frac 12 \, b \, T_c^{-\frac 12} \,  (T \!- \! T_c) + \mathcal{O}[(T \!- \!T_c)^2]$, with $T_c$ given by Eq.~(\ref{eq:TNeel}). More formally, Eq.~(\ref{eq:Invchi}) is analytical $\forall \, T \! > \! 0$. Hence, its explicit $\sqrt{T}$ behavior, driven by the presence of Kohn lines on the FS, gets reflected by a critical exponent $\gamma = \frac 12$ {\sl only} at quantum criticality, when $T_c \!=\!0$ and a Taylor expansion for $\chi_m^{-1}$ is no longer possible.

On this basis, we are now able to consistently identify the different regions in the $T\!-\!n$ phase-diagram around the QCP illustrated in Fig.~\ref{fig:Fig1}(c). Starting from the left side for $n\! > \! n_c$, a region with incommensurate AF (``i-AF'') 
order \cite{Rampon2025} is found below the critical temperature of Eq.~(\ref{eq:TNeel}), which displays the above-mentioned {\sl parabolic} behavior in $n \! - \! n_c$. Right above $T_c$, the classical critical regime [gray area marked with ``C" in Fig.~\ref{fig:Fig1}(c)] is identified as the parameter region, where the lowest (linear) order of the Taylor expansion of $\chi_m^{-1}$
in $T \! - \! T_c$ dominates over the next one, featuring classical mean-field critical exponents, e.g., $\chi_m^{-1} \propto (T \! - \! T_c)^{\gamma_\text{cl}}$ with $\gamma_\text{cl} \! = \! 1$. By decreasing the density, for $n \lesssim n_c$, the quantum critical [``QC'', red shadowed area in Fig.~\ref{fig:Fig1}(c)] and the quantum disordered (``QD'', in green) regimes can be respectively identified as the regions where the Kohn-line driven quantum critical behavior in Eq.~(\ref{eq:Invchi}), i.e.,~$\chi_m^{-1} \propto \sqrt{T}$  (i.e., $\gamma \! = \!  \frac 12$), dominates the $T$-independent term $\!\propto\!(n_c\! -\! n)$ or vice versa. Hence, $b \sqrt{T} \gg a |n_c -n|$ defines the QC regime, while  $a(n_c -n) \geq b \sqrt{T}$ the QD one. We note here that, while the precise location of the borders delimiting the QC and QD regimes, due to their intrinsic crossover nature, depend, in general, on the quantitative criterion chosen \footnote{The criterion used is the dominance of the second w.r.t.~the first term of the r.h.s.~of Eq.~(\ref{eq:Invchi2}). Specifically, in Fig.~\ref{fig:Fig1}, we used the following relations: $T= \frac{16a^2}{b^2}(n_c -n)^2$,  $T= \frac{a^2}{b^2}(n_c -n)^2$ with $n < n_c$, and  $ T= \frac{1.8a^2}{b^2}(n_c -n)^2$, respectively, for the QC, QD and C regimes.}, their overall {\sl parabolic} shape in the $T\!-\!n$-phase diagram is directly dictated by the explicit expression in Eq.~(\ref{eq:Invchi}) and represents, thus, a genuine feature of our DMFT results. 

\begin{figure}[t!]
  \centering
  \hspace{-0.45cm} \includegraphics[scale=0.92]{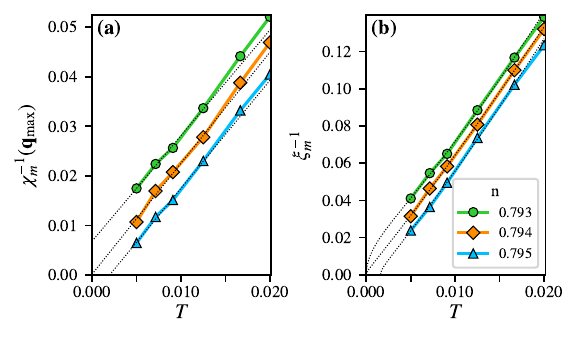}
  \vspace{-0.9cm}
\caption{(a) $T-$dependence of the inverse of the maximum magnetic susceptibility $\chi_m^{-1}(\mathbf{q}_\text{max})$ and (b) of the corresponding magnetic correlation length $\xi_m^{-1}$  for the DMFT solution of the 3D Hubbard model on a frustrated cubic lattice ($t'= 0.2$)  for different values of the electronic density $n$. The fits with Eq.~(\ref{eq:Invchi2}) and Eq.~(\ref{eq:Invxi2}) and $\sigma=1$ are shown with dotted lines.}
\label{fig:Fig2new}
\vspace{-5mm}
\end{figure}

{\sl Correlation length.} Differently from  $\chi_m$, the numerical data for the magnetic correlation length $\xi_m$ shown in Fig.~\ref{fig:Fig1}(b) do not allow for an immediate understanding. 
As we will see, $\xi_m$ actually entails the pivotal information to clarify the actual origin of the observed deviations of the quantum critical behavior from the HMM prediction.

Let us start by explicitly analyzing the effect of the Kohn anomalies on the O.Z. expression  Eq.~(\ref{eq:OZfit}) in the $T\rightarrow 0$ limit.
In particular, a FS featuring lines of KPs causes the emergence of non-analytical momentum dependencies in the non-interacting (static) magnetic susceptibility at $T=0$ \cite{Rice1970,Schaefer2017,Suppl}, i.e.,~$\chi^0(\pi,\pi, q_z) \! \sim \!\chi^0_{\rm max} - C |\delta q_z|^\sigma$, with $\delta q_z \! = \! (q_z \! - \! \bar{q_z})$ and $\sigma \! = \! \frac 12$ (up to log-corrections).
For the DMFT case of a momentum-independent irreducible vertex in the (magnetic) BSE (\ref{eq:BSE}), such non-analytical features get transferred  to the interacting case \footnote{As noted in \cite{Vilardi2018}, this can be partly ascribed to the \emph{locality} of the irreducible vertex of DMFT, and typically  holds, for the magnetic sector, in the  correlated \emph{metallic} regime.}, leading 
to a breakdown of the O.Z. behavior of $\chi_m$ for $T\rightarrow 0$. In fact, the fitting of the coefficient ${\cal A}^{-1}$ in Eq.~(\ref{eq:OZfit}) yields the same singular behavior $\propto T^{-(2-\sigma)} = T^{-\frac 32}$  we derive analytically in \cite{Suppl} at the level of RPA. By  evaluating Eq.~(\ref{eq:OZfit}) for $q_z=\bar{q}_z$, one gets \cite{Suppl}:
\vspace{-1mm}
\begin{equation}
\xi_m^{-1}(T,n) = \sqrt{{\cal A} \; \chi_m^{-1}({\bf q}_\text{max})} \! \propto \! \sqrt{T^{\frac 32}\left[a(n_c \! - \! n) + b\sqrt{T}\right]}.
\label{eq:Invxi}
\end{equation}
This expression has been indeed used to obtain accurate fits of the data in Fig.~\ref{fig:Fig1}(b), explaining not only the approximate $\!\propto\!T$ behavior of $\xi_m^{-1}$ at high $T$ but also its low-$T$ regime. In particular, the validity of Eq.~(\ref{eq:Invxi}) has significant implications in the $T \rightarrow 0$ limit.
In fact, at the classical phase-transition for $n \!> \! n_c $, the property of ${\cal A}$ of being nonzero at $T\!=\!T_c$ and the possibility of Taylor expanding $\chi_m^{-1}$ for $T \! \gtrsim T_c \! > \!0$ yield a standard mean-field critical behavior for $\xi_m^{-1} \! \propto \! \sqrt{T \!- \!T_c}$, i.e.~$ \nu_{cl} \! = \! \frac 12$. At the same time, for $n \leq n_c$,  ${\cal A}$  \emph{vanishes} in the $T\! =\! 0$-limit, which corresponds to a \emph{divergent} spin-stiffness in the ground-state of the system (cf.~Appendix A in End Matter). 

Hence, it becomes clear how such a Kohn-anomaly-driven feature introduces a \emph{second}, independent diverging scale in the problem, forcing $\xi_m(T \! \rightarrow \! 0)$ to diverge not only at the QCP, but also in the \emph{whole} QD regime, where $\chi_m(T \! \rightarrow \! 0)$ is finite. Such a dichotomic  $T\!=\!0$-behavior of $\chi_m$ and $\xi_m$ in the QD regime is indeed a \emph{direct hallmark} of the Kohn-anomalies physics.

More quantitatively, for $T \!  \rightarrow \! 0$, Eq.~(\ref{eq:Invchi}) gives: (i) $\xi_m^{-1} \! \propto  \! T$ for $n \!= \! n_c$ and (ii) $\xi_m^{-1} \! \propto \! T^\frac 34$ for $n \! < \! n_c$, cf.~inset in Fig.~\ref{fig:Fig1}(b). Evidently, (i) features a non-HMM value of the corresponding quantum critical exponent ($\nu \! = \!1$) and the observed ``inversion'' of the Fisher relation. Conversely, (ii) reflects the general lack of a well-defined exponential decay of the i-AF spatial correlations in the $T\!=\!0$ limit. Indeed, due to the Kohn anomalies, the $T\!=\!0$ correlations display a \emph{power law} long-distance behavior. By Fourier transforming \cite{Gelfand1964}  the ground-state magnetic response, one find a \emph{long-ranged/power law} decay of the magnetic correlations \footnote{For instance, in the QD region, one has: $\chi_m(q_z) \! \sim \mbox{const.} \,  - |\delta q_z|^{\sigma} \propto (1 + \frac{|\delta q_z|^{\sigma}}{\rm const.})^{-1}$
(with $\sigma \!= \! \frac 12$), so that the long-distance magnetic correlations scale generally as $\chi_m(r) \! \sim \! \frac{1}{r^{d+\sigma}} \!  \equiv \! \frac{1}{r^{3.5}}$}.
In particular, at the
QCP, where the mass of the magnetic propagator vanishes, one finds  $\chi_m(q_z) \! \sim \!  |\delta q_z|^{-\sigma}$ corresponding to $\chi_m(r) \! \sim \! \frac{1}{r^{d-\sigma}} \!  \equiv \! \frac{1}{r^{2.5}}$. 
This $T\!=\!0$ effect, \emph{entirely} caused by the Kohn anomalies, evidently hinders a conventional definition \footnote{At $T \!= \! 0$, following the convention adopted for the LR case, it could be formally possible to define an $\eta_{LR} = 2 - \sigma$, which is non-zero also for MF theories. This LR-anomalous exponent would be however not associated to the scale of an exponential decay, but rather to the crossover scale $\xi_{LR}$ at which $\frac{1}{r^{2.5}}$ sets in, and, as such, is not suited to define the long-range decay at any finite $T$, including the QC regime of our interest.} of the non-thermal quantum critical exponent $\nu_n$ \footnote{As for the exponent $\nu_n'$ on the ordered phase, even if this was definable, its calculation would be extremely hard, due to the several non-trivial phase-transitions possibly occurring in the i-AF regime at low $T$ \cite{Rampon2025,Scholle2024}}. At the same time, it unveils how \emph{quantum} phase-transitions in our electronic system, despite the \emph{on-site} interaction, can acquire \emph{some} features of the universality class of systems with \emph{long-range interactions} \cite{Fisher1973} (namely with $U(r) \! \sim \! \frac{1}{r^{d+\sigma}}$). A quite relevant one is, arguably, the corresponding scaling relation, which would read  $\gamma \!  =  \! \sigma \, \nu  \!  = \! \frac{1}{2} \, \nu$, consistent with our results of $\gamma \! = \! \frac 12, \nu \! = \! 1$ (s.~Appendix B).


\begin{figure*}[t!]
  \centering
  \hspace{-1cm}
  \includegraphics[scale=0.8]{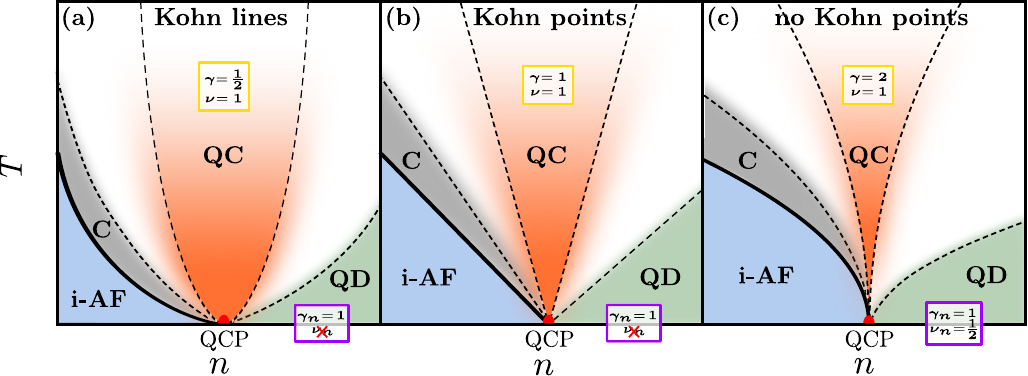}
 \vspace{-1mm}
  \caption{Sketch of the phase-diagrams expected, at the DMFT level, in the proximity of the i-AF QCP of a single orbital Hubbard model in 3D, whose FS entails (a) lines of Kohn points, (b) isolated Kohn points, (c) no Kohn points at all. 
}
\label{fig:Fig3}
\end{figure*}

{\sl Generalization to other FS geometries.} The physical insight gained from the interpretation of our results for (a) lines of Kohn points allows us to outline their generalization to the other two relevant cases for 3D single-orbital systems, i.e., to the systems, whose FS only displays (b) isolated Kohn points or (c) no Kohn points at all. In a nutshell, by adapting our derivations (see \cite{Suppl} for more details) to these FS geometries \cite{Schaefer2017,Stepanenko2017}, we find that the low-$T$ behavior of magnetic susceptibility in Eq.~(\ref{eq:Invchi}) is generalized as follows:
\begin{equation}
\chi^{-1}_m({\bf q}_{\rm max}, T,n)= a(n_c -n) + bT^\sigma,
\label{eq:Invchi2}
\end{equation}
with $\sigma\!=\!1$ for (b), and $\sigma\!=\!2$ for (c), in contrast to the previous (a) case where $\sigma\! = \! \frac 12$. To prove the validity of Eq.~(\ref{eq:Invchi2}) for the case (b) we report in Fig.~\ref{fig:Fig2new}(a) our DMFT calculations for $\chi_m$ of a 3D Hubbard model with the same $U$, but on a cubic lattice with geometric frustration \footnote{Specifically, the following 3D-dispersion with $t' =0.2t$ has been considered:
$\epsilon_{\mathbf{k}} = -2t\sum_{i=1}^3\cos k_i
-4t'\sum_{i<j}\cos k_i\cos k_j
$.}, whose FS features isolated KPs \cite{Fus2024PA}: The linear $T-$dependence we observe perfectly matches  Eq.~(\ref{eq:Invchi2}) predictions with $\sigma\!=\!1$.
Finally, in the absence of KPs, the $T-$dependence of $\chi_m$ for a correlated metal reduces to a Sommerfeld expansions, yielding trivially $\sigma\!=\!2$ and reflecting a full restoration of the conventional O.Z. description for $T \! \rightarrow \! 0$.

Similar to Eq.~(\ref{eq:Invchi}), Eq.~(\ref{eq:Invchi2}) does not only control the exponents in the QC region (where $\gamma\!=\! \sigma$), but also the overall structure of the associated phase diagram see Fig.~\ref{fig:Fig3}. 
Specifically, the $n$-dependence of $T_c = \left[\frac{a(n-n_c)}{b}\right]^\frac{1}{\sigma}$ at the QCP qualitatively differs in the presence of (a) lines of Kohn points, (b) isolated Kohn points  and (c) in their total absence, with an (a) {\sl quadratic},  (b) {\sl linear} or (c) {\sl square-root} behavior of the critical temperature in the $T\!-\!n$ phase-diagram, respectively. Evidently, the same will happen to our estimates for the crossover borders of both QC and QD regimes, respectively, just as before but for another $\sigma$, yielding the qualitative phase diagrams sketched in Fig.~{\ref{fig:Fig3}.

At the same time, the QC behavior of $\xi_m$ remains affected by an (albeit milder) breakdown of the O.Z. expression for $T \rightarrow 0$  only in the case of isolated KPs, as ${\cal A}(T \rightarrow 0) \propto T$ for (b) and ${\cal A}(T \rightarrow 0) = {\cal A}_0 >0 $ for (c), i.e. altogether ${\cal A}(T \! \rightarrow \! 0) \! \propto \! T^{2-\sigma}$.
Hence, we have, in general:
\begin{equation}
\xi_m^{-1}(T,n)   \propto \! \sqrt{T^{2-\sigma}\left[a(n_c \! - \! n) + bT^\sigma\right]},
\label{eq:Invxi2}
\end{equation}
whose validity for case (b), i.e.,~$\sigma\!=\!1$, is explicitly demonstrated, via our DMFT calculations, in Fig.~\ref{fig:Fig2new}(b).

As for the exponents, according to Eqs.~(\ref{eq:Invchi2})-(\ref{eq:Invxi2}), we have $\nu \!= \! 1$ in all cases, featuring a milder modification of the HMM Fisher relation in the QC regime for (b), with $\gamma \!= \! \nu= \! 1$, and its full restoration in (c). For the same reason, while Eq.~(\ref{eq:Invchi2}) implies $\gamma_n \!= \!1$, the non-thermal exponent $\nu_n$ can be defined only for (c), with $\nu_n \! = \! \frac{\gamma_n}{2} \! = \! \frac 12$. 

{\sl Spatial correlation effects.} As DMFT is a MF theory in space, it is worth discussing how 
its results can be modified by the spatial correlations of 3D systems. 
These surely affect \cite{Dare1996,Antipov2014,Hirschmeier2015,DelRe2019}} the 
 exponents of the finite-$T$ transition, which  must become $\gamma_{cl} \! \simeq \! 1.39, \,  \nu_{cl} \! \simeq \! 0.71$ of the 3D Heisenberg class \cite{Zinn1996,Holm1993,Garioud2024}. Arguably, also the $T_c$ dependence on $n$ computed in DMFT \footnote{We note that the unusual {\sl quadratic/linear} dependencies of $T_c$ on $n_c \! - \! n$, obtained in 3D within DMFT, may also emerge in mean-field calculations \cite{Scholle2023} for 2D systems with correspondingly analogous FS geometries \cite{Scholle2024}.} may be modified.
 On the other hand, significant changes of our results in the QC/QD regimes can only be expected if the FS of the DMFT metallic state (cf.~App.~C) gets destroyed by spatial correlations, which is unlikely in 3D for the $n$ values considered.
More specifically, for $d_\text{eff} \! > \! 4$, the non-thermal QC exponents will be unaffected, i.e., $\gamma_n \! = \! 1$ and, if no KPs are present, $\nu_n \! = \! \frac 12$. 
It is harder to foresee, whether 
nonlocal correlations in 3D systems with Kohn anomalies can overturn the 
Kohn-driven effects on thermal QC exponents ($\gamma,\nu$) already captured by DMFT, calling for future investigations.
If no KPs are present, instead, the HMM values ($\gamma \!= \! 2\nu \! = \! \frac 32$ \cite{Loehneysen2007}) should be recovered.
 

{\sl Conclusions.} We illustrated the general features of magnetic quantum criticality in correlated metals by means of DMFT and rigorous analytical derivations, unveiling the impact of Kohn anomalies on the quantum critical exponents as well as on {\sl whole} 
phase-diagram around the associated QCP. In this context, we were able to nail down the precise mechanism breaking the conventional HMM description: The Kohn anomalies on the FS drive a \emph{power-law} decay of the magnetic correlations in the ground state of the system, modifying the universal properties of the corresponding \emph{quantum} phase-transition \emph{both} at zero and at finite $T$. Hence, the FS geometry should be explicitly considered for classifying the universal behavior of quantum phase-transitions in correlated metals.


This is particularly relevant for interpreting QC features emerging in {\sl ab-initio} many-body calculations (often performed with DFT+DMFT schemes \cite{Kotliar2006,Held2007}), since the effects we described, especially the isolated KP case, occur for relatively common cubic lattice symmetries \cite{Stepanenko2017}. Furthermore, our results represent a non-trivial starting point for studying quantum phase-transitions in layered/lower-dimensional \cite{Roy2008,Klein2020,Debbeler2023,Lunts2023} correlated quantum materials, as well as for systems, in which the coupling of the electronic systems to phononic degrees of freedom is expected to be significant \cite{Rodiere2026}.


\begin{acknowledgments}
\textit{Acknowledgements.} 
D.R.F., S.A., K.H. and A.T.~acknowledge support from the Austrian Science Fund (FWF), Grant-DOI:10.55776/KIN2563725 (project P1 of the research unit QUAST, for5249, of the German Research Foundation, DFG), K.H. through 
SFB Q-M\&S (Grant-DOI: 10.55776/F86), and A.T. through Grant-DOI: 10.55776/I5487. T.S. acknowledges support from the German Research Foundation, DFG, through DFG ID No.~449872909 (project P4 of the research unit QUAST, for5249). A.V. acknowledges the hospitality of the Coll{\`e}ge de France, Paris. Note: Our international collaboration (Austria-Germany-Russia) on this specific topic started {\sl before} February 2022, hence not violating the current regulatory prescriptions. The authors thank Federico Becca, Massimo Capone, Patrick Chalupa-Gantner, Sergio Caprara, Nicol\`o Defenu, Samuele Giuli,  Motoharu Kitatani, Emin Moghadas, Walter Metzner, Andrew J. Millis, Georg Rohringer, Michael Scherer, Robin Scholle, Florian Streuselberger, Qimiao Si, Carlo Tortora, and Demetrio Vilardi for fruitful discussions. The authors gratefully acknowledge the Austrian Scientific Computing (ASC), where part of the calculations were performed, as well as scientific support and HPC resources provided by the computing service facility of the MPI-FKF.
\end{acknowledgments}


\bibliography{Vertex}

\clearpage 
\onecolumngrid 
\setcounter{page}{1}
\setcounter{secnumdepth}{1}

\begin{center}
    \textbf{\large End Matter}\\[0.5em]
\end{center}

\thispagestyle{empty}
\twocolumngrid

\noindent
\textit{Appendix A: Kohn-anomalies driven divergence of the spin stiffness: The Fisher relation in the quantum critical regime.} 

\noindent
At a first glance, the observed modification of quantum critical exponents and of their Fisher relation (s.~main text) for an electronic system with \emph{on-site} interaction and with Kohn anomalies on the FS may appear a somewhat surprising result.

Let us recall
the standard conditions \cite{Fisher1964,Kenna2014}, under which the Fisher relation is derived, are (i) the fluctuation-dissipation theorem, and (ii) the existence of a single length scale, whose $T \! \rightarrow \! 0$ divergence at QCP is uniquely related to the occurrence of the quantum phase-transition considered. As we discuss below, in the QC regime the condition (ii) is directly affected by the presence of Kohn points (KPs) on the FS.


In general, quantum critical exponents may differ from their HMM or mean-field (MF) values due to non-analyticities in the momentum dependence of the magnetic susceptibility. For instance, at low dimensions, these non-analyticities can be generated by critical fluctuations. This mechanism is fundamentally different from the Kohn-driven non-analyticities studied here, which originate from the Fermi-surface geometry and exist \emph{already} away from criticality. Since critical fluctuation-induced non-analyticities do not invalidate assumptions (i) and (ii) above, the Fisher relation remains valid: $\gamma=(2-\eta)\nu$ (or $\gamma \! = \! 2\nu$ for $d_{\mathrm{eff}}\! = \! d+z \! > \! 4$, as in our case). Thus, for the i-AF–PM quantum phase transition in conventional 3D metals without Kohn anomalies, the standard HMM exponents, $\gamma=3/2$, $\nu=3/4$, and $\eta=0$, are recovered, satisfying $\gamma=2\nu$.

To highlight the difference w.r.t.~the Kohn-anomalies case discussed later, it is useful to present a concise derivation of the Fisher relation in our context. 
By assuming that \emph{no other} possible sources of non-analyticities than those intrinsically associated to the (quantum) phase-transitions exist, the expression for the (inverse) magnetic susceptibility for small $\delta q_z = q_z - \bar{q}_z$ around its maximum [e.g., located at $(\pi, \pi, \bar{q}_z)$], would read:
\begin{equation} \begin{array}{cc}\chi_m^{-1}(q_z,T)  \simeq  & \hspace{-25mm} \chi_{\rm max}^{-1}(T) + {\cal D}_S(T) (\delta q_z)^{2-\eta} \\   & \hspace{-5mm} \propto    \chi_{\rm max}^{-1}(T) \left[ 1 + \left({\cal D}_S(T) \chi_{\rm max}(T) \right) (\delta q_z)^{2-\eta} \right]. \end{array} \label{eq:EM1}  \end{equation}
Here,  $\chi_{\rm max}(T) = \chi_m(\bar{q_z}, T)$, while the coefficient ${\cal D}_S(T)$, which would correspond to the  ${\cal  A}^{-1}(T)$ in our O.Z. notation [cf.~Eq.~(\ref{eq:OZfit}) in the main text], is
often referred to as \emph{``spin stiffness''}.
It is then evident that, at (quantum) criticality, the corresponding length $\xi_m$ will scale as 
\begin{equation} 
\xi_m^{2-\eta} = \left[{\cal D}_S(T) \chi_{\rm max}(T) \right] 
\Rightarrow   \xi_m^{-1} \! \propto \left[{\cal D}_S^{-1}(T) \, \chi_{\rm max}^{-1}(T) \right]^\frac{1}{2-\eta}.
\label{eq:EM2} \end{equation}
In the conventional case, where there are no other sources of non-analyticity than those associated to the criticality itself (fully encoded in the exponent $\eta$), the spin stiffness will be an analytical, positive definite function of the temperature, i.e.~${\cal D}_S(T) \simeq {\cal D}^0_S + {\cal O}(T)$, where ${\cal D}^0_S$ is a finite, positive constant. Hence, the divergence of the susceptibility as $\chi_{\rm max}^{-1}(T) \propto T^{ \gamma}$ at the quantum phase-transition will automatically imply a divergence of the corresponding correlation length as \begin{equation}
    \xi_m^{-1} \propto T^\nu \propto \left[({\cal D}^0_S)^{-1} \, T^\gamma \right]^\frac{1}{2-\eta}, \label{eq:EM3} \end{equation}
fulfilling the Fisher relation: $\gamma = (2-\eta) \nu$. Note that, for $d_{\rm eff} > d_c$ (our case), one has $\eta =0$, reducing Eq.~(\ref{eq:EM3}) to $\gamma = 2 \nu$. Hence, from now on, we will set $\eta \equiv 0$.

The situation is considerably different in the presence of Kohn anomalies on the FS. As shown in Refs.~\cite{Rice1970,Schaefer2017,Stepanenko2017,Suppl}, their occurrence triggers a \emph{non-analytical} momentum-dependence of the \emph{non-interacting} magnetic susceptibility at $T \! = \! 0$ around its maximum, namely, as $\sim |\delta q_z|^{\sigma}$ with $\sigma \! = \! \frac 12$ for lines of KPs and $\sigma \! = \! 1$ for isolated KPs.
Due to the metallic nature of the ground state of the 3D HM at the considered densities (cf.~Appendix C), the Kohn-driven non-analyticities of the bubble term around $\bar{q}_z$ get directly transferred in the fully \emph{interacting} magnetic susceptibility of DMFT \cite{Vilardi2018}. 
However, while the non-interacting bubble term  does not display any divergence (and hence any phase-transition), the inclusion of local vertex corrections of DMFT does allow, 
instead, for a divergence of $\chi_m(\bar{q_z}, T\!= \!0)$ at a given density $n_c$. 

Evidently, \emph{this} quantum phase-transition will be now affected by the non-analytical momentum-dependencies associated to the Kohn anomalies.
In fact, the Kohn-driven non-analyticities directly control the long-distance spatial decay of the corresponding magnetic fluctuations at $T \! = \! 0$.
Specifically, if  $\chi_m(q_z) \! \sim const. \,  - |\delta q_z|^{\sigma}$ (as in the QD region) or  $\sim |\delta q_z|^{-\sigma}$ (at the QCP) with $0 \!< \! \sigma \! < \! 2$, in the \emph{long-wavelength} limit the $|\delta q_z|^{\sigma}$ term will eventually prevail  over the conventional $(\delta q_z)^2$. 
This leads to a breakdown of the O.Z. expression [Eq.~(\ref{eq:OZfit})] in the ground-state, featuring an intrinsically \emph{long-range} (LR) power-law behavior of the magnetic correlations, namely \cite{Gelfand1964}: $\chi_m(r) \! \sim \! \frac{1}{r^{d+\sigma}}$ in the QD regime and $\chi_m(r) \! \sim \! \frac{1}{r^{d-\sigma}}$ at the QCP.
Evidently, a modified LR dependence of the magnetic correlations has a direct impact on the universal behavior of the corresponding quantum phase-transition at $T\! = \!0$, since the predominance in the RG scaling of the above-discussed $|\delta q_z|^\sigma$-term (with $0 \!< \! \sigma \! < \! 2$) over the O.Z.-term $|\delta q_z|^2$ directly modifies the corresponding fixed-point at the QCP. This feature, which occurs independently of the RG-scaling of the Hubbard interaction term and directly modifies the critical field-theoretical description of our \emph{quantum} phase-transition, is, thus, not ascribable to the (possible) presence of dangerously irrelevant variables \footnote{Possible effects, e.g., on the \emph{finite-size} scaling estimates of the critical exponents for the Ising model in high-dimensions have been systematically analyzed in \cite{Kenna2014,Flores2016}.} for $d \ge d_c =4$ . 

At finite $T$, the universal properties of the classical transition are obviously not affected by this modification, since the Kohn-driven non-analytic effects are smoothed out by the Fermi-Dirac distribution for $T \! > \! 0$. Quite remarkably, however, the \emph{thermal} quantum critical exponents in the QC funnel above the QCP are instead fully subjected to the Kohn-driven changes, despite the O.Z.-based Eq.~(\ref{eq:EM2}) still holds at finite $T$.


The reason thereof can be illustrated by explicitly discussing the modifications to Eq.~(\ref{eq:EM2}) due to the Kohn anomalies: The KP-driven non-analytical momentum dependence of $\chi_m$ at $T=0$ induces two \emph{concomitant} effects: (i) the maximal susceptibility at $n = n_c$ scales as $\chi_m^{-1}(\bar{q_z},T) \equiv \chi_{\rm max}^{-1}(T) \sim T^{\, \sigma}$; (ii) the associated spin stiffness
${\cal D}_S(T)$ of the corresponding O.Z. expression will become \emph{singular} at $T=0$. In particular, the spin-stiffness  diverges as ${\cal D}_S(T) \sim {\cal A}^{-1}(T) \sim T^{\, -(2- \sigma)}$ with $0 <\sigma <2$ (cf.~$T-$dependencies of ${\cal A}^{-1}(T)$ given in the main text).
Hence, Eq.~(\ref{eq:EM2}) now yields: 
\begin{equation}
\label{eq:nu_Kohn}
\xi_m^{-1}(T) \propto T^\nu \sim \left[\frac{\chi_{\rm max}^{-1}(T)}{{\cal D}_S(T)}\right]^{\frac
12} \! \propto \left[\frac{T^{\sigma}}{T^{\sigma-2}} \right]^{\frac 12} = \,  T \Rightarrow \,
\boxed{\nu = 1} \, ,
\end{equation}
while
\begin{equation}
\chi_{\max}^{-1}(T) \propto T^{\, \gamma} \sim T^{\, \sigma}
\; \Longrightarrow \;
\boxed{ \phantom{\frac 12} \gamma = \sigma \equiv \sigma\,\nu \phantom{\frac 12}} \; , 
\label{eq:Fisher-Kohn}
\end{equation}
where, in the last step, we used the fact that $\nu \! \equiv \!1$ for any value of $0 \! < \! \sigma  \! < \!  2$, i.e.,~for all relevant Kohn-driven cases which we considered.
Comparing Eqs.~(\ref{eq:EM2})-(\ref{eq:EM3}) to Eqs.~(\ref{eq:nu_Kohn})-(\ref{eq:Fisher-Kohn}) highlights the \emph{distinct nature} of the two kinds of non-analyticity discussed above: These are encoded, respectively, (i) into the anomalous critical exponent $\eta$ [for those originating from the (quantum) critical fluctuations in low dimensions] and (ii) into the exponent $0 \!< \! \sigma \!< \!2$, for the emergent long-range correlation features driven by the Kohn-anomalies.\\

\noindent {\sl Appendix B: Relation with the LR universality class.} 

\noindent
The modified expression for the Fisher relation can be rationalized by explicitly considering, for a moment, the universality class of systems with LR interactions \cite{Fisher1973}. In fact, for (Ising) spin systems with LR interaction of the kind of $U(r) \sim \! \frac{1}{r^{d+\sigma}}$ (with  $0 \! < \! \sigma  \! < \! 2$) the corresponding magnetic correlations decay as $\chi_m(r) \sim \! \frac{1}{r^{d-\sigma}}$  i.e.,~\emph{exactly} as the $T=0$ spatial correlations at our QCP.
For this LR-interaction class, the corresponding (Fisher) relation between $\gamma$ and $\nu$ reads $\gamma = \sigma \nu$ \cite{Fisher1973}, consistent with the results for our Kohn-driven QC case. However,  the origin of the LR effects in the two cases is \emph{different} and, therefore, the analogy is \emph{not} complete: The modified Fisher relation for LR interactions is valid also for the classical (finite T) transition, while the Kohn-driven modifications can \emph{only} affect the quantum critical exponents (and relations thereof), representing an emerging properties of the ground-state correlations. 
For the same reason, the values of the Kohn-driven exponents in the QC regime ($\gamma \! = \! \sigma, \nu \! \equiv \! 1$) are \emph{not} necessarily the same as in the LR universality class (i.e., in $d=3$, $\gamma=1, \nu= \frac{1}{\sigma}$).
This illustrates the necessity of explicitly including the Kohn-anomaly properties of the FS for a proper classification of the universal properties of \emph{quantum} magnetic transitions in bulk correlated metals.\\

\noindent {\sl Appendix C: One-particle properties in the quantum critical region of DMFT.}

\begin{figure}[b]
\vspace{-5mm}
  \centering
  \hspace{-1cm} \includegraphics[scale=0.75]{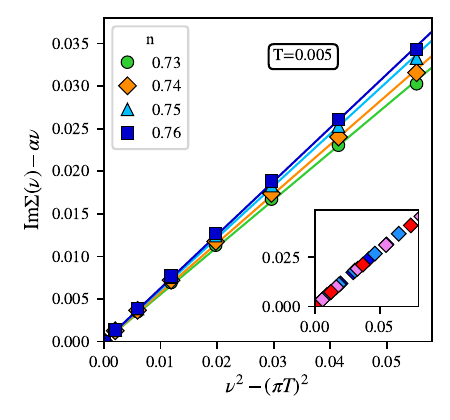}
  \vspace{-0.5cm}
  \caption{ Imaginary part of $\Sigma(\nu)$ computed in DMFT at $T=0.005$ at different densities, after the subtraction of its linear part $\alpha\nu$, as a function of $\nu^2 - (\pi T)^2$, i.e. of the expected frequency dependence for a Fermi liquid system \cite{Maslov2012,Chubukov2012}. Inset: Corresponding data for $n\!=\!n_c$ at different $T$, i.e.~$0.005$ (blue), $0.0055$ (light blue), $0.0062$ (violet) and $0.0125$ (red), all displaying the same  \textbf{$\nu^2 - (\pi T)^2$}.}
  \label{fig:Fig2}
\end{figure}

\noindent
The description of our DMFT results in the main text can be completed by the study of one-particle spectral properties.  In Fig.~\ref{fig:Fig2}, we analyze the Matsubara frequency $i\nu$ behavior of the (local) DMFT self-energy $\Sigma_\nu$. In particular, the Fermi liquid nature of all datasets considered (including the one at $n \! = \! n_c \simeq 0.74$) is demonstrated by the dependence of the scattering part of Im~\!$\Sigma_\nu$ on $\nu^2 - (\pi T)^2$ \cite{Chubukov2012,Maslov2012}, obtained after subtracting the quasiparticle mass-renormalization term $\alpha \nu$ from Im~\!$\Sigma_\nu$ (which contributes to Re~\!$\Sigma$ for real frequencies). This reflects the lack of feedback of nonlocal correlations on the one-particle spectral properties in the DMFT scheme.

Nonlocal corrections to DMFT in $\Sigma$ may alter the Fermi liquid behavior of Fig.~\ref{fig:Fig2}, if additional Kohn-anomalies-driven corrections \cite{Holder2014,Sykora2018,Sykora2021} arise.\\

\end{document}